\documentstyle[12pt,aaspp4,flushrt]{article}

\begin{document}

\title{The Second Cambridge Pulsar Survey at 81.5\,MHz}
\author{J. A. Shrauner\altaffilmark{1}, J. H. Taylor\altaffilmark{2}}
\affil{Joseph Henry Laboratories and Physics Department \\
       Princeton University, Princeton, NJ 08544}
\author{G. Woan\altaffilmark{3}}
\affil{Mullard Radio Astronomy Observatory, Cavendish Laboratory\\
       Cambridge University, Cambridge CB3 OHE, UK}
\altaffiltext{1}{Present address: WebTV Network, 1295 Charleston Road,
  Building B, Mountain View, CA~~94043; shrauner@corp.webtv.net}

\altaffiltext{2}{joe@pulsar.princeton.edu} \altaffiltext{3}{Present
  address: Department of Physics \& Astronomy, University of Glasgow, 
  Glasgow G12 8QQ, UK; graham@astro.gla.ac.uk}

\begin{abstract}
We have searched the northern sky for pulsars at the low radio
frequency of 81.5\,MHz, using the 3.6-hectare array at Cambridge,
England.  The survey covered most of the sky north of declination
$-20^\circ$ and provided sensitivities of order 200\,mJy for pulsars
not too close to the galactic plane.  A total of 20 pulsars were
detected, all of them previously known.  The effective post-detection
sampling rate was 1.3\,kHz, and the sensitivity to low-dispersion
millisecond pulsars was sufficient to allow the detection of objects
similar to PSR J0437$-$4715 (period 5.7\,ms, dispersion measure
2.6\,cm$^{-3}$pc, mean flux density $\sim1\,$Jy).  No such pulsars
were found.

\end{abstract}


\section{Introduction}

Pulsars were discovered in 1967 when Bell and Hewish were observing
the interplanetary scintillations (IPS) of compact radio sources,
using a fixed dipole array at 81.5\,MHz (\cite{hbp+68}).  Chart
recordings were examined by eye for evidence of intrinsically pulsed
signals, which looked somewhat different from the scintillating
quasars being studied.  Within a few months, the first Cambridge pulsar
search discovered a total of six pulsars: PSRs B0329+54, B0809+74,
B0834+06, B0950+08, B1133+16, and B1919+21
(\cite{hbp+68,phbc68,cp68}).  Soon afterward the dipole array returned
to nearly full-time studies of IPS, compact radio sources, and the
interplanetary medium---the purposes for which it had been designed.

Most pulsar searching since 1968 has been done at radio frequencies
around 400\,MHz and higher, where the sky background is lower,
interstellar dispersion and scattering are less of an impediment, and
much larger bandwidths can be gainfully used.  Nevertheless, in 1993
it seemed to us for a number of reasons that a second Cambridge pulsar
search at 81.5\,MHz would be a worthwhile undertaking.  The IPS dipole
array has been doubled in size since 1968, and now has a geometric
collecting area of 36,055\,m$^2$ or 3.6 hectares---approximately half
that of the 305\,m telescope in Arecibo, Puerto Rico.  The discovery
of PSR~J0437$-$4715 by Johnston {\em et al.} (1993) \nocite{jlh+93}
demonstrated the existence of at least one very strong, nearby,
millisecond pulsar.  We reasoned that if similar pulsars exist in the
northern sky, the 3.6-hectare array should be capable of detecting
them, especially if their low-frequency radio spectra are steeper than
those of most slowly rotating pulsars, as suggested by Foster {\it et
al.}\ (1991).\nocite{ffb91}  Fortunately, the nature of the Cambridge
array is such that co-opting any one of its 16 simultaneous beams for
pulsar searching can be done with only minimal disruption to its
continuing IPS studies.

\section{The Telescope}

The 3.6 hectare array is a meridian-transit instrument consisting of
4096 full-wave dipoles operating at a wavelength $\lambda=3.68\,$m.
The dipoles are arranged in 32 east-west rows of 128 dipoles each,
with a spacing between rows of $d=0.65$ wavelengths.  A branched
feeder network combines the signals within each row, properly phasing
them for a celestial source on the meridian.  Summed signals from the
16 northern-most and 16 southern-most rows are then combined in a
matched pair of ``Butler matrices'' to form simultaneous beams at 16
declinations in the range $-8^\circ<\delta<90^\circ$.  For IPS studies
the two half-arrays are used as a north-south phase-switching
interferometer (\cite{ptr+87}); however, for our pulsar observations
we connected the two halves as a total-power phased array.  A modified
set of 16 beams can be generated by inserting an extra phase gradient
in the north-south direction across the entire array, thereby shifting
each beam north by half a beamwidth, or about $3^\circ$.

The declination of the peak response for each of the 32 possible beams
is given by
\begin{equation}
  \label{eqn:camb_d0}
  \delta_0(N) = 52.16^\circ + 
    \arcsin\left[{\frac{\lambda (N-10)}{16d}}\right] \,,
\end{equation}
where the beam number, $N$, is an integer in the range 1--16 for the
unshifted beams, and a half-integral value between 1.5 and 16.5 for
the shifted beams (\cite{pur81,ptr+87}).  The beams have half-power
widths of $5.5^\circ/\cos{[52.16^\circ-\delta_0(N)]}$ in declination.  A
reflecting screen lies $\lambda /4$ beneath each row, and is
inclined towards the south at 50 degrees to the vertical in order to
increase the sensitivity of the array at lower declinations.  However,
Tappin (1984) \nocite{tap84} has shown that the declination power
response, $D$, of the antenna follows that expected for an array of
dipoles $\lambda /4$ above a flat, horizontal reflecting screen; he
finds no evidence for increased sensitivity at lower declinations.

\begin{figure}
  \begin{center}
	\plotone{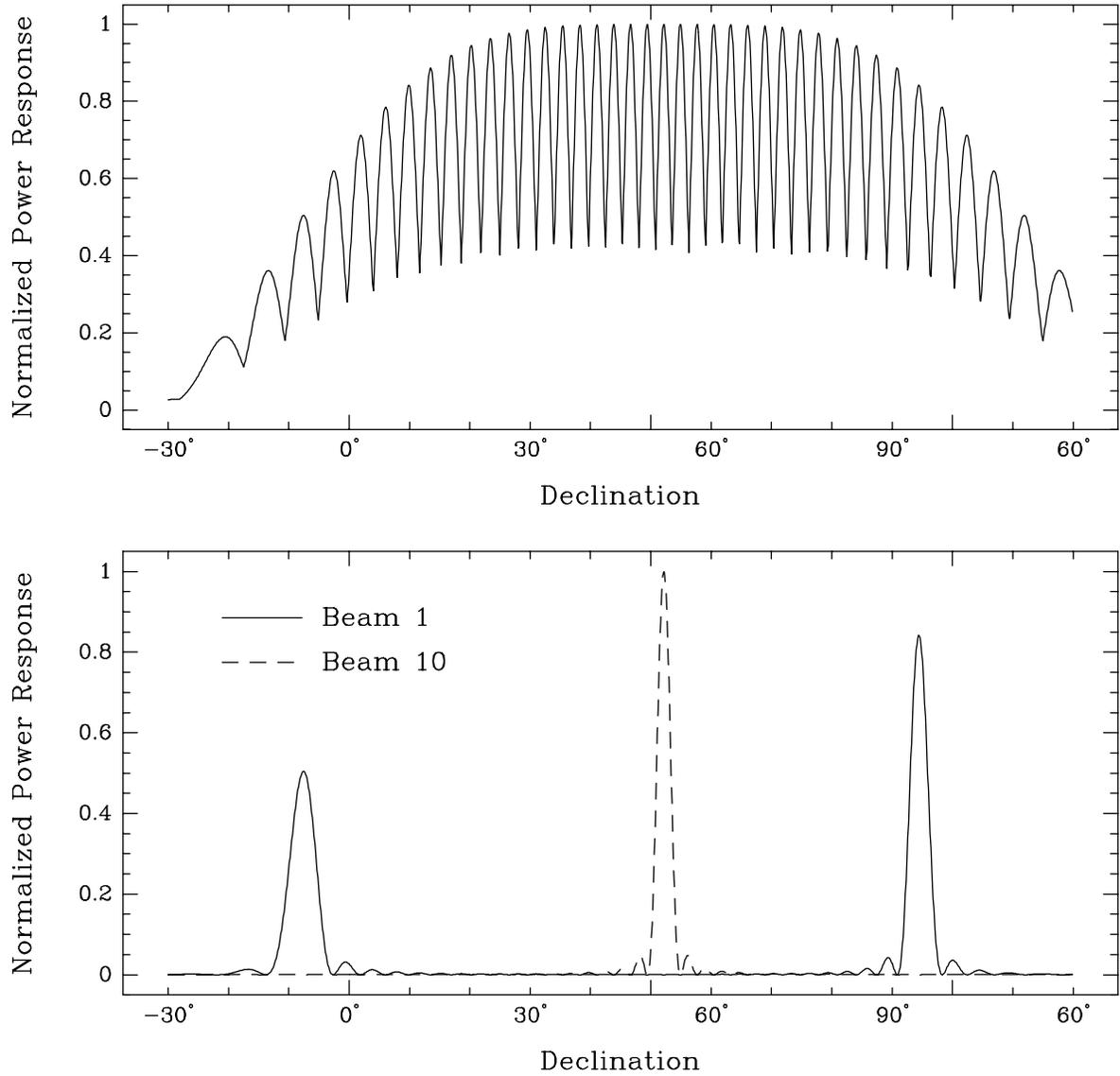}
    \caption{Normalized declination response of the 3.6-hectare array.
    {\em Top:} combined power response provided by all 32 beams.  {\em
    Bottom:} full response patterns of beams 1 and 10.  Note that beam
    1 has a secondary response at $+94^\circ$ which is stronger than its
    primary response at $-8^\circ$.}  \label{fig:camb_beams}
  \end{center}
\end{figure}

Let $\eta$ represent the angle of the reflecting screen to the
horizontal, and define the following three additional angles:
\begin{eqnarray}
  \phi   &\equiv& 52.16^\circ - \delta, \\*
\rule[-5.5mm]{0mm}{12.5mm}
  \psi   &\equiv& \left(\frac{\pi d}{\lambda}\right)\sin{\phi} + 
                \left(\frac{N-10}{16}\right)\pi, \\*
  \alpha &\equiv& \left(\frac{32\pi d}{\lambda}\right)\sin{\phi}.
\end{eqnarray}
The normalized declination power response can then be expressed as
\begin{equation}
  \label{eqn:camb_D}
  D\left(N,\delta\right) = \left(\frac{1\pm\cos{\alpha}}{2}\right)\,
     \sin^2\left[{\frac{\pi}{2}\cos\left({\phi-\eta}\right)}\right]\,
     \frac{\sin^2{16\psi}}{{16^2}\sin^2{\psi}},
\end{equation}
where the three factors on the right-hand side arise from the
interference between the two array halves, the reflecting screen, and
the 16 phased rows in each half, respectively.  Following Tappin, we
take $\eta=0$.  Inserting the peak-response declinations of
equation~(\ref{eqn:camb_d0}) into equation~(\ref{eqn:camb_D}) then
shows that $D$ is maximized by phasing the array so as to use the plus
sign in the first factor for integral beam numbers, and the minus sign
for half-integral beam numbers.  The top panel of
Figure~\ref{fig:camb_beams} shows the combined declination response
afforded by the antenna for the whole survey, obtained by summing the
calculated power responses of all 32 beams.

The peak gain of each beam is proportional to the effective collecting
area of the array, which falls off as the cosine of zenith angle.
Because of this foreshortening, several of the beams have secondary
responses comparable to or even greater than their primary responses.
As examples, the bottom panel of Figure~\ref{fig:camb_beams} shows the
individual power responses of beams 1 (solid curve) and 10 (dashed
curve).  The true declinations of sources detected in the beams with
large secondary responses can be determined from their transit times
through the east-west beam: a source at declination $\delta$ has a
half-power transit time of ${107}/\!\cos{\delta}$ seconds
(\cite{pur81}).  For the secondary responses at $\delta>90^\circ$,
transit times are shifted by 12 hours.

\section{Observations and Data Analysis}

Our observations were carried out between November 1993 and June 1994.
We phased the array for a given beam number, $N=1$, 1.5, 2, 2.5,
\dots, 16.5, and recorded data continuously for 24 hours as the sky
drifted overhead.  Observations were repeated as demanded by
interference, equipment malfunction or mis-adjustment, etc., until each
of the 32 beams had been observed and produced high-quality data at
least twice.  Figure~\ref{fig:camb_skymap} illustrates the full sky
coverage of the survey.  We analyzed beams 16 and 16.5 for signals
entering via their secondary responses, rather than the primary ones.
Consequently the survey provided useful sensitivity between
declinations of $-20^\circ$ and $+86^\circ$, with a few small areas
missing because of interference.  Note, however, that because of array
foreshortening at low declinations and the ``scalloped'' overall
response as a function of $\delta$, as well as the large variation in
background temperature over the sky, our sensitivity even to
low-dispersion, long period pulsars varied by more than an order of
magnitude over the surveyed area.  As described in more detail below,
the median sensitivity for low-dispersion pulsars with periods
$P>0.1\,$s was about 200\,mJy.

When the 3.6-hectare array is being used solely for IPS observations,
the southernmost 14 beams are observed simultaneously.  We diverted a
single beam on a given day for pulsar observations, thereby causing
only minimal disruption to the IPS work.  Signals from the two halves
of the array, properly phased for the desired beam, were extracted
from the two Butler matrices and added.  The resulting signal was
mixed to intermediate frequency of 10.7\,MHz and then to baseband,
using quadrature local oscillators.  An automatic gain control served
to keep the post-detection noise level constant.  The in-phase
(``real'') and quadrature (``imaginary'') baseband signals were
low-pass filtered at 0.47\,MHz (the filters are 60\,dB down at
0.5\,MHz), sampled at a 1\,MHz rate, and digitized in a 12-bit
analog-to-digital converter controlled by a programmable digital
signal processor (DSP; Analog Devices ADSP-21020) on a VME board made
by the Ixthos Corporation.

The incoming data were double-buffered in fast memory on the DSP
board.  At intervals of 256\,$\mu$s the ADSP-21020 performed a
256-point complex Fourier transform on the incoming data and squared
the resulting magnitudes.  Three consecutive power spectra were
summed, and every 768\,$\mu$s the difference between the 256-channel
power spectrum and an exponentially-weighted mean of spectra over the
past 1.5\,s was one-bit sampled.  The resulting bits were packed into
32-bit words and passed to a controlling workstation, which wrote them
onto magnetic tape along with the running averages and
root-mean-square deviations of the summed power spectra and the date,
time, and beam number.  A full day's observing session generated about
3.5\,GB of data, which fit comfortably on a single 8\,mm tape
cassette.

Searching for pulsar signals was accomplished by reading the recorded
data from tape into another workstation.  The one-bit data were
unpacked and divided into consecutive blocks with length comparable to
the beam transit time at the relevant declination.  Block lengths of
$2^{17}$ samples were used for beams 1--10, 16, and 16.5; $2^{18}$
samples for beams 10.5--13; and $2^{19}$ samples for beams 13.5--15.5.
These numbers correspond to transit times of 101, 201, and 403\,s.
Beams 16 and 16.5 were processed using the short block length so as to
remain sensitive to pulsars in their secondary responses at
declinations near $-20^\circ$ and $-13^\circ$, respectively.  The
total amount of computing was substantial: with the DEC AXP\,3000/400
workstation used to do most of it, about 3.6 days of computing was
required for each day of observations---nearly eight months of
continuous computing, in all.

The search program analyzed each beam area independently.  Progressive
delays were introduced between the 256 spectral channels in a manner
that optimally produces time series at different dispersion measures
(\cite{tay74}).  Curvature in the delay-versus-frequency relation was
compensated by shifting the middle two-thirds of the spectral channels
by one channel.  For each data block, the one-bit samples were
optimally de-dispersed at 256 dispersion measures evenly spaced
between 0 and 12.78\,cm$^{-3}$pc.  Additional dispersion ranges were
generated by first adding successive pairs of time samples in each
spectral channel, and then reapplying the de-dispersion algorithm.
The de-dispersed time series in each new range thus had half the
length of those in the previous range, and the upper half of each
group of dispersions extended the ranges of those already computed.  In
this manner, each data block was also de-dispersed by 128 dispersion
measures evenly spaced across each of the following ranges:
12.83--25.60, 25.65--51.25, and 51.30--102.55\,cm$^{-3}$pc.

For each data block a total of 640 de-dispersed time series were
Fourier transformed and analyzed for harmonic content.  After removing
known sources of periodic interference (in particular, the 274\,Hz
phase-switching frequency used for IPS studies), the search program
identified the strongest periodicities by interpolating between the
complex Fourier coefficients and finding the peak amplitudes, and also
by summing harmonically related frequencies in groups of 2, 3, 4, 8,
and 16 harmonics.  The program isolated the most promising pulsar
suspects in the frequency domain, formed the equivalent integrated
pulse shapes by transforming complex coefficients (with up to 32
harmonics) back into the time domain, and computed their
signal-to-noise ratios.  Any suspect detections with signal strengths
above a predetermined threshold were presented as possible detections.

\section{Flux Density Measurements and Survey Sensitivity}

The expected signal-to-noise ratio, ${\cal R}$, of a received pulsar
signal can be related to various system parameters by the expression
\begin{equation}  \label{eqn:camb_snr}
  {\cal R} = \frac{\eta_Q\,S_{81}\,G}{T_{\rm sys}}
   \left[\frac{(B \tau n_p)(P-w)}{w}\right]^{1/2}\,,
\end{equation}
where $\eta_Q$ is the digitization efficiency factor, $S_{81}$ the
mean flux density of the pulsar at 81.5\,MHz (Jy), $G$ the telescope
gain (K~Jy$^{-1}$), $T_{\rm sys}$ the system temperature (K), $B$ the
bandwidth (Hz), $\tau$ the integration time (s), $n_p$ the number of
polarizations, $P$ the period, and $w$ the pulse width.  Our system
had $B = 1$\,MHz, $n_p = 1$, and because of the one-bit post-detection
digitization, $\eta_Q = \sqrt{2/\pi}$ (\cite{dtws85,tms86}).  Flux
density measurements can therefore be obtained for detected pulsars by
determining $G/T_{\rm sys}$ from calibration measurements, calculating
${\cal R}$, $P$, and $w$ from average profiles, and solving for
$S_{81}$.

The appropriate value of $G$ for any direction in our surveyed region
depends on variations in antenna gain across the declination strip,
the peak gain and sidelobe responses of the different beams, and the
background sky temperature.  The first effect can be calculated from
the declination power response given by equation~(\ref{eqn:camb_D}),
and the latter two by using a sky map of equivalent background flux
density obtained with the same telescope.  Figure~4.5.2 of Purvis
(1981) presents a contour map of the system noise level of the
3.6-hectare array when operated as a phase-switching interferometer.
The contour levels indicate twice the receiver noise level, in
Janskys, for a bandwidth $B=1$\,MHz and integration time $\tau_{\rm
map}=0.33$\,s.  The map therefore plots the flux density $S_{\rm map}$
at which unresolved, continuous sources at the beam centers would be
detected with signal-to-noise ratio ${\cal R} = 2$.  At an arbitrary
declination $\delta$ close to the strip-center at $\delta_0$, we have
\begin{equation}
  \frac{G}{T_{\rm sys}} =
   \frac{2(B \tau_{\rm map})^{1/2}}{S_{\rm map}}\,
   \frac{D(N,\delta)}{D[N,\delta_0(N)]}\,.
\end{equation}
It therefore follows that the flux density of a pulsar observed with
signal-to-noise ratio $\cal R$ is given by 
\begin{equation}
  \label{eqn:camb_flux}
  S = {\cal R}\,S_{\rm map}\ 
     \frac{D[N,\delta_0(N)]}{D(N,\delta)}\,
     \left[\frac{0.13 w}{\tau(P-w)}\right]^{1/2}.
\end{equation}

More than $10^8$ possible combinations of period, dispersion measure,
and pulse width are tested for significance in each data block, and
consequently a moderately high threshold signal-to-noise ratio must be
used to keep the rate of false detections low.  We adopted a lower
limit of ${\cal R}=8.5$, and the sensitivity of the whole survey can
therefore be characterized by inserting this number into equation
(\ref{eqn:camb_flux}) together with appropriate values for the
integration time, $\tau$, and effective pulse width, $w$, including
all forms of instrumental smoothing.  The pulse width is bounded on
the low side by $\Delta t$, the post-detection sample interval; it
also depends on the dispersion measure.  For periods smaller than
$64\Delta t \approx49.2$\,ms in the lowest dispersion range, the
minimum detectable flux density scales approximately as $P^{-1/2}$.
Similar behavior obtains for the higher dispersion ranges, with the
effective $\Delta t$ doubling in each successive range.  A more
complete sensitivity analysis based on the detailed implementation of
our search algorithm leads to the stepwise continuous sensitivity
curves plotted in Figure~\ref{fig:camb_minflux} (see \cite{nft95}).
This graph shows the sensitivity of our survey to a low-duty-cycle
($w\ll P$) pulsar in a region with the median sky background noise,
$S_{\rm map} = 2$\,Jy.  The curves show the minimum flux densities
that yield ${\cal R}=8.5$, as a function of period, and are plotted
for five values of DM up to the largest values searched.  Steps in the
sensitivity curves occur at periods equal to multiples of $\Delta t$,
where the number of harmonics tested by the search software changes.
The sky background noise $S_{\rm map}$ varies between the extremes of
0.5 and 9\,Jy, but is $<2$\,Jy for 50\% and $<3$\,Jy for about 70\% of
the visible sky (\cite{pur81}).  It is important to note that the
sensitivity models of equation~(\ref{eqn:camb_flux}) and
Figure~\ref{fig:camb_minflux} do not explicitly account for
interstellar scattering effects, which will likely dominate the pulse
broadening for pulsars with ${\rm DM}\gtrsim50$\,cm$^{-3}$pc.  Such
pulsars always lie close to the galactic plane, and for this and other
reasons are more profitably observed at frequencies much higher than
81.5\,MHz.

\section{Results}

\begin{figure}
  \begin{center} 
    \plotone{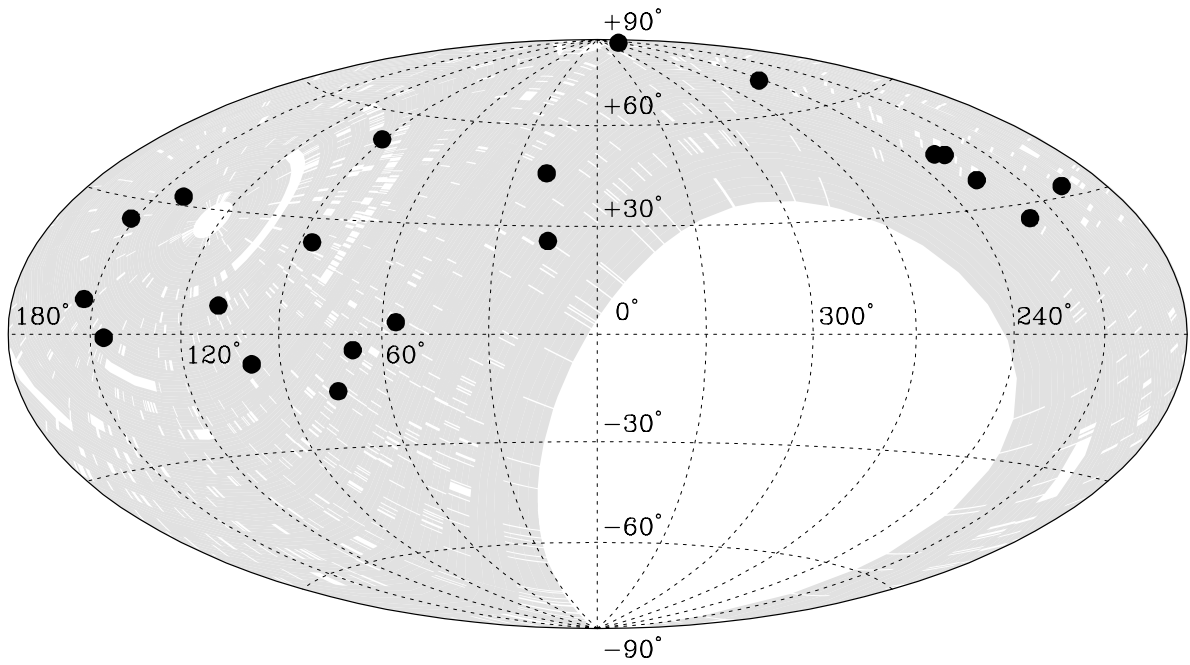}
    \caption{Shaded areas indicate sky coverage of the survey in
    galactic coordinates.  Small gaps are the result of local radio
    interference, and filled circles represent the 20 known pulsars
    detected in the survey.}  \label{fig:camb_skymap} 
  \end{center}
\end{figure}

\begin{figure}
  \begin{center}
    \plotone{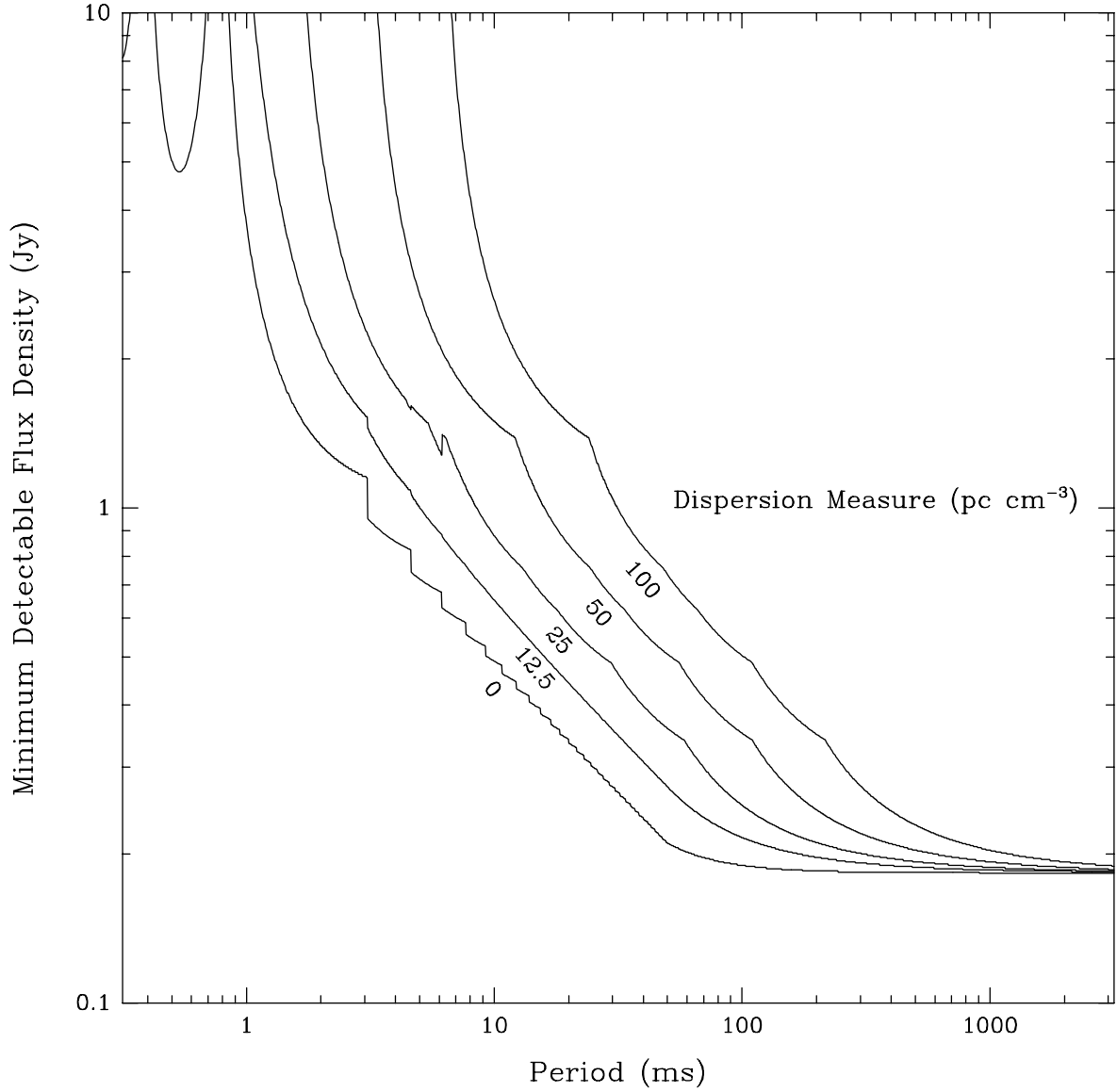}
    \caption{Minimum detectable flux density for low duty-cycle
      pulsars, plotted as a function of period for dispersion measures
      0, 12.5, 25, 50, and 100\,cm$^{-3}$pc.  At the higher
      dispersions, interstellar scattering will degrade sensitivities
      further.}
    \label{fig:camb_minflux}
  \end{center}
\end{figure}

\begin{figure}[f]
  \begin{center}
    \plotone{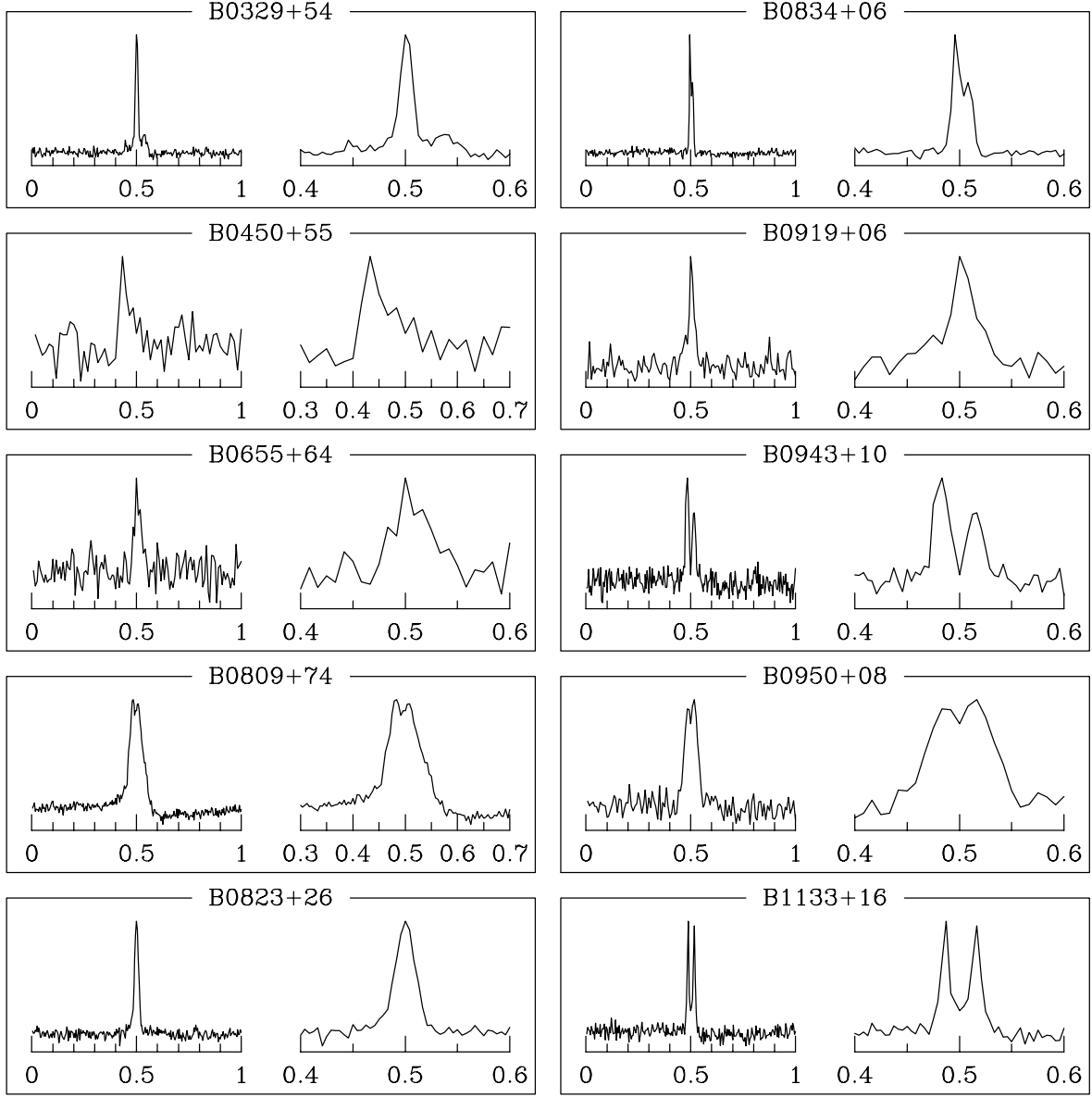}
    \caption{Pulse profiles for the 20 detected pulsars, each from a
      single transit through the telescope beam.  Intensity is shown
      over the entire period (left) and over a smaller region centered
      on the pulse.}
    \label{fig:camb_profs}
  \end{center}
\end{figure}

We detected a total of 20 pulsars, all of them previously known.
Their locations are plotted in galactic coordinates in
Figure~\ref{fig:camb_skymap}.  Table~1 lists their catalogued
parameters including names, B1950 celestial coordinates, periods,
dispersion measures, and 400\,MHz flux densities, $S_{400}$.  We also
list the offsets, $\Delta\delta$, between each pulsar's location and
the nearest beam center, as well as the observed signal-to-noise
ratios for the two times each pulsar was nominally observed, and the
average flux densities, $S_{81}$, for the positive detections.  The
signal-to-noise ratios are those determined by the search program for
the closest beam pointing.  We include values for both observations of
the relevant declination strip to give some indication of the
efficiency of the search algorithm and the repeatability of the
measurements.  Flux densities were calculated by applying
equation~(\ref{eqn:camb_flux}) to the average profiles created by
synchronously averaging one full beam-transit of data, centered on the
pulsar's right ascension.  The profile obtained in this way for each
pulsar is illustrated in Figure~\ref{fig:camb_profs}.  

The accuracy of our flux density measurements depends on calibration
errors and manual ``readout uncertainties'' of the sky map, as well as
the degree to which the pulsar signals were affected by time-variable
scintillation effects.  The sky map is thought to be accurate to
within about 20\%, and the map-reading uncertainties probably
contribute another 20\% uncertainty.  Interstellar scintillation
causes significant modulations of the signal strengths by factors of 2
or more, particularly for the lowest-dispersion pulsars.  Some
information on internal consistency of the results may be gained from
the 13 pulsars detected in each of the two days their declination
strip was observed.  Their flux density measurements show a
root-mean-square deviation of 25\% around the mean.  Altogether we
estimate that the flux density measurements are accurate to
approximately 40\%, and within such limits are in good accord with
previous work.  An especially low measured flux for PSR~B0809+74 may
be the result of uncertain extrapolation of sky-background contours
near the edge of the map.

Radio-frequency spectra of the 20 detected pulsars are presented in
Figure~\ref{fig:camb_spectra}.  Pulsar spectra are generally
characterized by power-law indices of $-1.5$ to $-2$ at frequencies
above a few hundred MHz; typically the spectra flatten and perhaps
even turn over at lower frequencies.  Low-frequency turnovers are
clearly evident for some of the spectra displayed in
Figure~\ref{fig:camb_spectra}, and it seems certain that similar
behavior in many other known pulsars prevented their detection in our
survey.  Even among the pulsars we detected, nearly half are an order
of magnitude weaker at 81.5\,MHz than would be expected from
straight-line extrapolations of their higher-frequency spectra.

Not surprisingly, all six of the original Cambridge pulsars were
detected in the survey, although PSR B0950+08 was one of the seven
pulsars only seen in a single day's observations.  None of the seven
pulsars seen only once would have been confirmed as a new pulsar
discovery solely on the basis of the observations we made; however,
further attempts at confirmation would have been undertaken for any
candidate with a single observation as promising as these.  PSR
B0950+08 was actually observed during initial trial runs using only
half the array, and also in some observations when the array was
mistakenly anti-phased.  Figure~\ref{fig:camb_spectra} shows
variations more than an order of magnitude in measured fluxes of
PSR~B0950+08 around 100\,MHz, with our measurement close to the top of
the range.  A scintillation-induced order-of-magnitude decrease in
received flux would have rendered the pulsar undetectable in the
survey.  The strongest pulsar detected in our survey, PSR~B2217+47,
was not detected in the first Cambridge pulsar search because, with a
${\rm DM}=44$\,cm$^{-3}$pc, its pulse is smeared out over more than a
full period in a 1\,MHz bandwidth.  Its pulses, and also those of PSRs
B0919+06 and B1642$-$03, are quite undetectable with the Cambridge
telescope, without the benefit of dispersion-removal techniques.

In the coldest regions of the sky the minimum detectable flux density
of our survey is about 100\,mJy.  Several pulsars we did not see have
been detected by others at frequencies between 70 and 100\,MHz, with
mean flux densities greater than that minimum: PSRs B0138+59,
B0320+39, B0523+11, B0525+21, B0531+21, B0611+22, B1612+07, B1842+14,
B2111+46, B2154+40, and B2319+60 (\cite{ikms81,sie73,mgj+94}).  Of
these, PSR~B2111+46 has a DM of 141.5\,cm$^{-3}$pc, which exceeds the
highest value we searched; and the pulsar in the Crab nebula, PSR
B0531+21, is undetectable as a periodic source because interstellar
scattering broadens its pulse excessively at 81.5\,MHz.  The remaining
pulsars all have reported flux measurements below the detection
threshold for their sky positions, periods, and dispersion measures,
except for PSRs B0138+59 and B0320+39.  Our estimated sensitivities at
these positions are 150 and 420\,mJy, respectively.  Izvekova {\em et
al.}  (1981) report a flux measurement for PSR B0138+59 of 260\,mJy at
85\,MHz.  Malofeev {\em et al.} (1994) quote fluxes for PSR B0320+39
of 700 and 500\,mJy at 60 and 100\,MHz, respectively, while Izvekova
{\em et al.}  (1981) give mean flux density measurements of 230 and
160\,mJy at the same frequencies.  It follows that these two pulsars
would have been marginal detections in our survey, at best.

In summary, the second Cambridge pulsar survey detected all previously
known pulsars in the northern hemisphere that we could have expected
to detect, with the possible exception of PSR B0138+59.  We obtained
the lowest frequency flux measurements available for PSRs~B0450+55,
B0655+64, B1839+56, B2110+27, and B2224+65.  Interestingly, PSR
B0655+64 is a binary pulsar, with orbital period just slightly longer
than a day.  If it had been just a little stronger, and therefore had
been discovered by Bell and Hewish in 1968, disentangling its orbiting
nature by means of observations with this transit telescope would have
been quite difficult --- as indeed it still was, when the pulsar was
discovered with the 300\,ft transit telescope in Green Bank, a decade
later (see \cite{dbtb82}).  With respect to possible detections of
millisecond pulsars, we offer the following comments.  Over the
northern celestial hemisphere our median sensitivity to pulsars
similar to J0437$-$4715 ($P=5.7$\,ms, DM$=2.6$\,cm$^{-3}$pc,
$S_{81}\approx1$\,Jy; see \cite{mabe96}) was approximately 0.7\,Jy.
If pulsars like PSR J0437$-$4715 were scattered all over the northern
hemisphere, we would have detected more than half of them.  We
conclude, therefore, that such objects are quite rare.  Almost
certainly there is no such pulsar in our surveyed region with
$S_{81}>3$\,Jy.

\acknowledgements We thank A.~Hewish and R.~Hills for organizational
assistance in the early phases of this project, and J.~Milrod and the
Ixthos Corporation for their generous loan of the DSP hardware.
Pulsar studies at Princeton are supported by the US National Science
Foundation.  G.~W. thanks SERC (now PPARC) for a Research Fellowship
and Clare College Cambridge for a College Lectureship, both held
during the bulk of this work.

\clearpage

%\bibliographystyle{apj1c}
%\bibliography{journals1,modrefs,psrrefs}

\begin{thebibliography}{}

\bibitem[Cole \& Pilkington 1968]{cp68}
Cole, T.~W. \& Pilkington, J. D.~H. 1968, Nature, 219, 574

\bibitem[Damashek {\it et al.}  1982]{dbtb82}
Damashek, M., Backus, P.~R., Taylor, J.~H., \& Burkhardt, R.~K. 1982, ApJ, 253,
  L57

\bibitem[Deshpande \& Radhakrishnan 1992]{dr92a}
Deshpande, A.~A. \& Radhakrishnan, V. 1992, J. Astrophys. Astr., 13, 151

\bibitem[Dewey {\it et al.}  1985]{dtws85}
Dewey, R.~J., Taylor, J.~H., Weisberg, J.~M., \& Stokes, G.~H. 1985, ApJ, 294,
  L25

\bibitem[Foster, Fairhead, \& Backer 1991]{ffb91}
Foster, R.~S., Fairhead, L., \& Backer, D.~C. 1991, ApJ, 378, 687

\bibitem[Gupta, Rickett, \& Coles 1993]{grc93}
Gupta, Y., Rickett, B.~J., \& Coles, W.~A. 1993, ApJ, 403, 183

\bibitem[Hewish {\it et al.}  1968]{hbp+68}
Hewish, A., Bell, S.~J., Pilkington, J. D.~H., Scott, P.~F., \& Collins, R.~A.
  1968, Nature, 217, 709

\bibitem[Izvekova {\it et al.}  1981]{ikms81}
Izvekova, V.~A., Kuz'min, A.~D., Malofeev, V.~M., \& Shitov, Y.~P. 1981,
  Ap\&SS, 78, 45

\bibitem[Johnston {\it et al.}  1993]{jlh+93}
Johnston, S. {\it et al.}  1993, Nature, 361, 613

\bibitem[Kuzmin {\it et al.}  1978]{kms+78}
Kuzmin, A.~D., Malofeev, V.~M., Shitov, Y.~P., Davies, J.~G., Lyne, A.~G., \&
  Rowson, B. 1978, MNRAS, 185, 441

\bibitem[Lorimer {\it et al.}  1995]{lylg95}
Lorimer, D.~R., Yates, J.~A., Lyne, A.~G., \& Gould, D.~M. 1995, MNRAS, 273,
  411

\bibitem[Lyne \& Rickett 1968]{lr68b}
Lyne, A.~G. \& Rickett, B.~J. 1968, Nature, 219, 1339

\bibitem[Malofeev {\it et al.}  1994]{mgj+94}
Malofeev, V.~M., Gil, J.~A., Jessner, A., Malov, I.~F., Seiradakis, J.~H.,
  Sieber, W., \& Wielebinski, R. 1994, A\&A, 285, 201

\bibitem[McConnell {\it et al.}  1996]{mabe96}
McConnell, D., Ables, J.~G., Bailes, M., \& Erickson, W.~C. 1996, MNRAS, 280,
  331

\bibitem[McLean 1973]{mcl73}
McLean, A. I.~O. 1973, MNRAS, 165, 133

\bibitem[Nice, Fruchter, \& Taylor 1995]{nft95}
Nice, D.~J., Fruchter, A.~S., \& Taylor, J.~H. 1995, ApJ, 449, 156

\bibitem[Pilkington {\it et al.}  1968]{phbc68}
Pilkington, J. D.~H., Hewish, A., Bell, S.~J., \& Cole, T.~W. 1968, Nature,
  218, 126

\bibitem[Purvis 1981]{pur81}
Purvis, A. 1981.
\newblock PhD thesis, Cambridge University

\bibitem[Purvis {\it et al.}  1987]{ptr+87}
Purvis, A., Tappin, S.~J., Rees, W.~G., Hewish, A., \& Duffett-Smith, P.~J.
  1987, MNRAS, 229, 589

\bibitem[Sieber 1973]{sie73}
Sieber, W. 1973, A\&A, 28, 237

\bibitem[Sieber \& Wielebinski 1987]{sw87}
Sieber, W. \& Wielebinski, R. 1987, A\&A, 177, 342

\bibitem[Slee \& Hill 1971]{sh71}
Slee, O.~B. \& Hill, E.~R. 1971, Aust. J. Phys., 24, 441

\bibitem[Stinebring \& Condon 1990]{sc90a}
Stinebring, D.~R. \& Condon, J.~J. 1990, ApJ, 352, 207

\bibitem[Tappin 1984]{tap84}
Tappin, S.~J. 1984.
\newblock PhD thesis, Cambridge University

\bibitem[Taylor 1974]{tay74}
Taylor, J.~H. 1974, A\&AS, 15, 367

\bibitem[Taylor, Manchester, \& Lyne 1993]{tml93}
Taylor, J.~H., Manchester, R.~N., \& Lyne, A.~G. 1993, ApJS, 88, 529

\bibitem[Taylor {\it et al.}  1995]{tmlc95}
Taylor, J.~H., Manchester, R.~N., Lyne, A.~G., \& Camilo, F. 1995, Unpublished
  (available at ftp://pulsar.princeton.edu/pub/catalog)

\bibitem[Thompson, Moran, \& Swenson 1986]{tms86}
Thompson, A.~R., Moran, J.~M., \& Swenson, G.~W. 1986, { {I}nterferometry and
  {S}ynthesis in {R}adio {A}stronomy}, (New York: John Wiley and Sons)

\end{thebibliography}

\addtocounter{figure}{-1}
\begin{figure}[f]
  \begin{center}
    \plotone{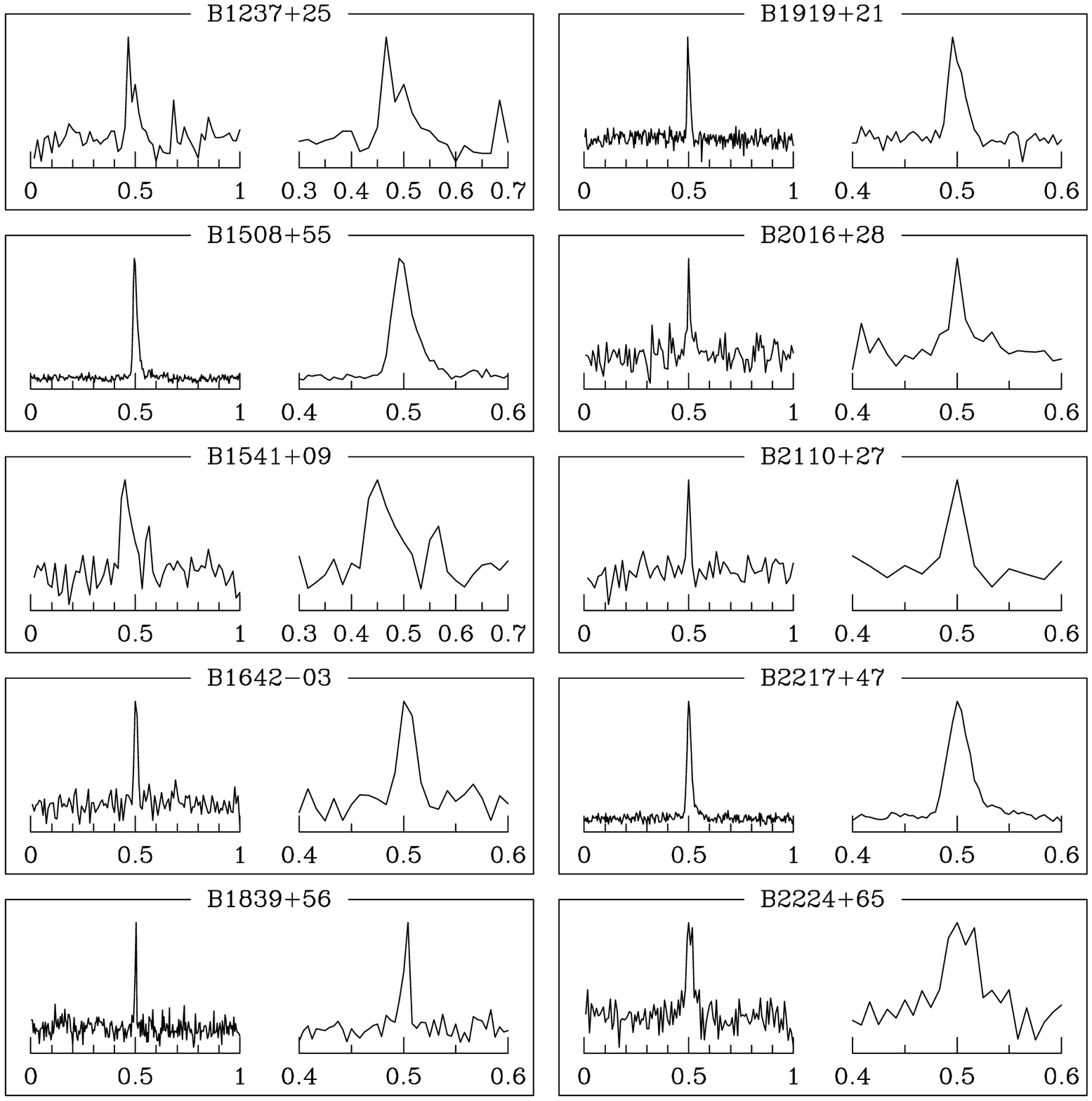}
    \caption{Pulse profiles, continued.}
  \end{center}
\end{figure}

\begin{figure}[tb]
  \begin{center}
    \plotfiddle{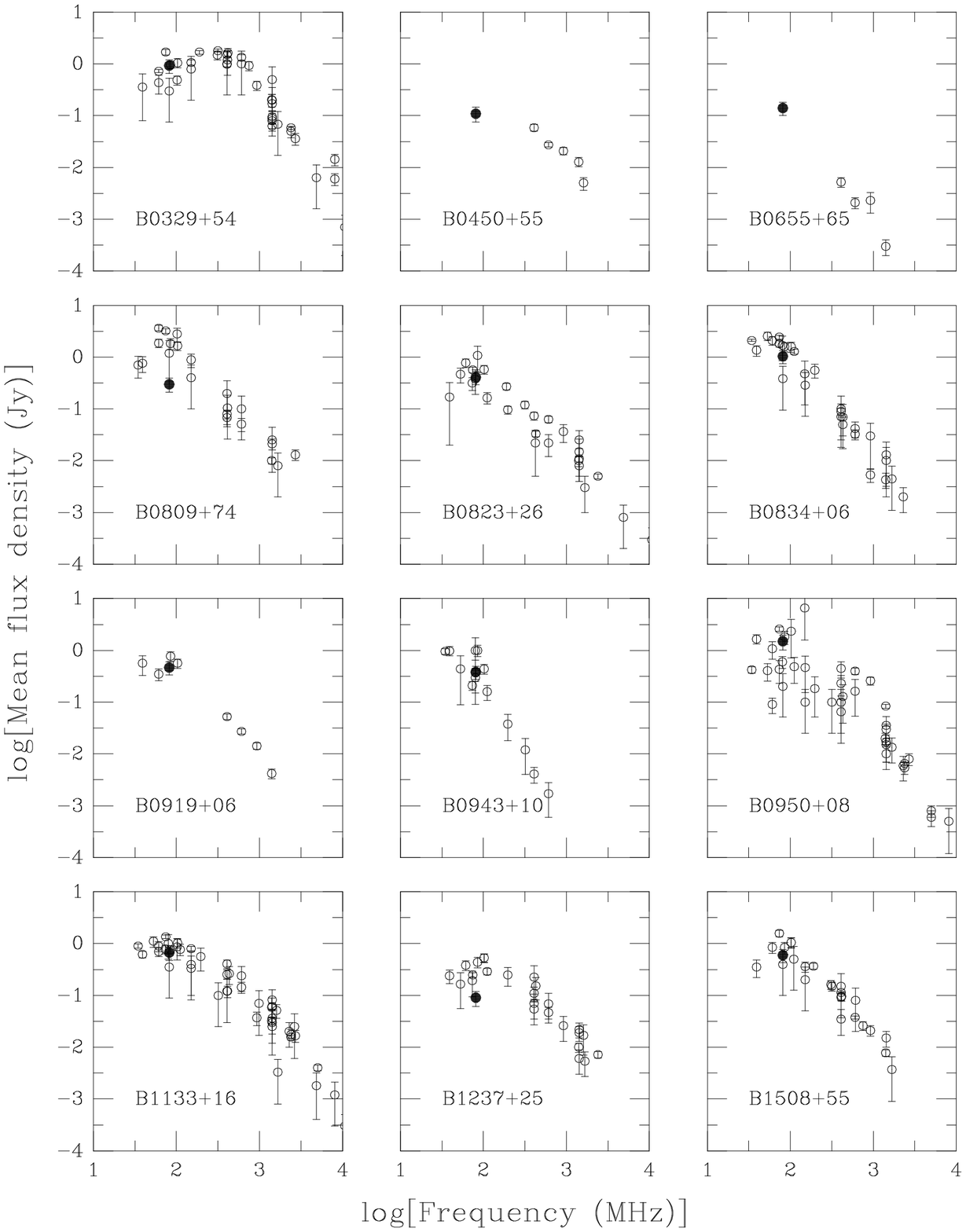}{6.0in}{0}{70}{70}{-220}{-40}
    \caption{Radio frequency spectra of the 20 detected pulsars.  Flux
      density measurements obtained in this survey are plotted as
      filled circles; those from other studies as open circles.
      Published results are taken from \protect\cite{sh71,lr68b,phbc68,lylg95,sie73,ikms81,grc93,dr92a,sw87,kms+78,sc90a,mcl73,tmlc95}.}
    \label{fig:camb_spectra}
  \end{center}
\end{figure}

\addtocounter{figure}{-1}
\begin{figure}[tb]
  \begin{center}
    \plotfiddle{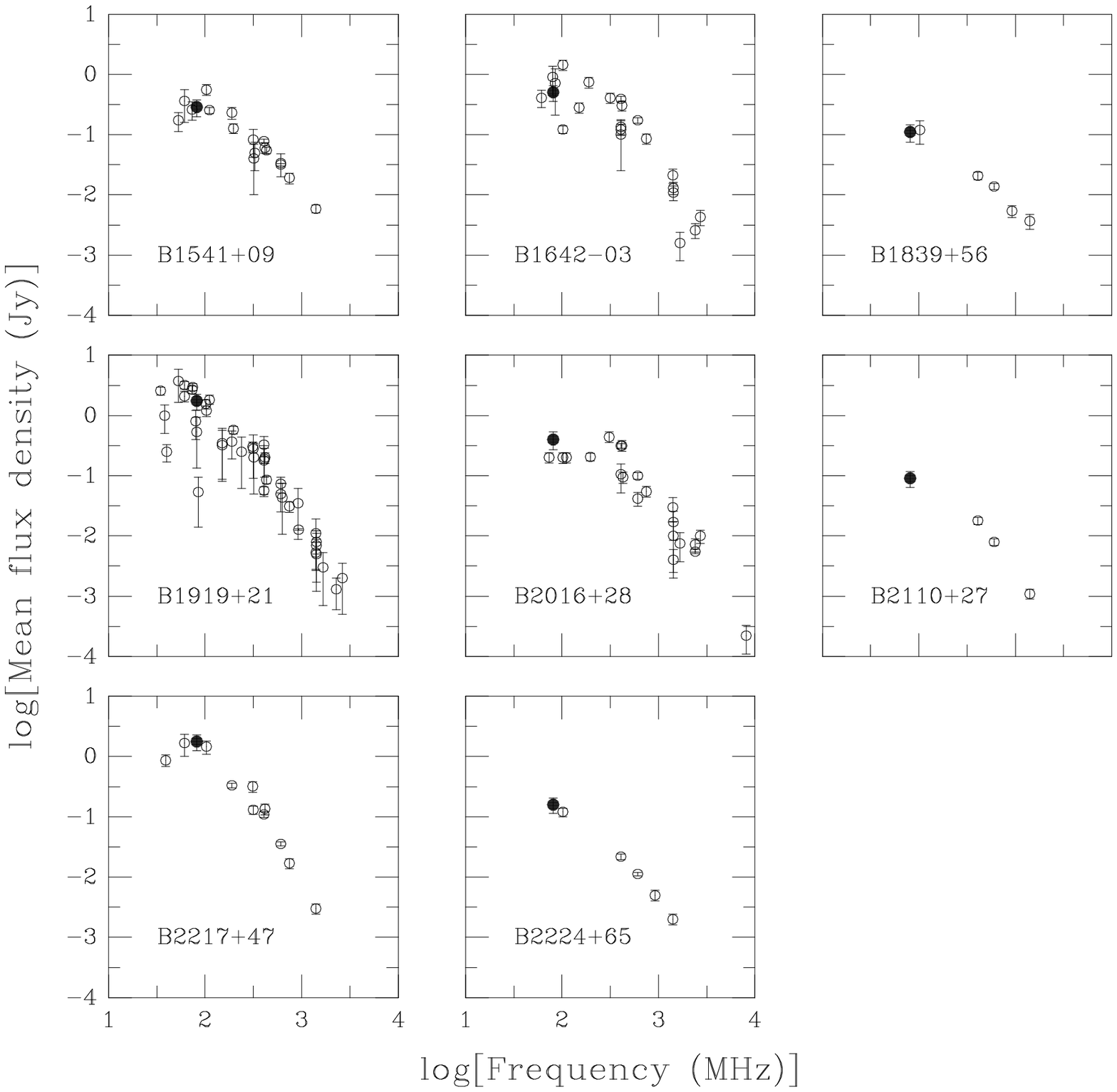}{6.0in}{0}{70}{70}{-220}{-40}
    \caption{Radio frequency spectra, continued.}
  \end{center}
\end{figure}

\clearpage
\begin{table}[tb]
  \caption{Known pulsars detected in the second Cambridge 
           survey.\tablenotemark{\star}} \label{tab:camb_oldpsrs}
  \begin{center}
    \leavevmode
    \small
     \begin{tabular}{lccccrrrrr}
      \hline 
      \hline
      \multicolumn{1}{c}{PSR}     &
      \multicolumn{1}{c}{R.~A.}    &
      \multicolumn{1}{c}{Dec.}     &
      \multicolumn{1}{c}{$P$}      &
      \multicolumn{1}{c}{DM}      &
      \multicolumn{1}{c}{$\Delta\delta$}  &
      \multicolumn{2}{c}{${\cal R}$}  &
      \multicolumn{1}{c}{$S_{400}$} &
      \multicolumn{1}{c}{$S_{81}$}    \\
      &
      \multicolumn{1}{c}{(B1950)}  &
      \multicolumn{1}{c}{(B1950)}  &
      \multicolumn{1}{c}{(s)}      &
      \multicolumn{1}{c}{(cm$^{-3}$pc)} &&
      &
      &
      \multicolumn{1}{c}{(mJy)}    &
      \multicolumn{1}{c}{(mJy)}    \\
      \hline
      B0329+54 &$03^{\rm h} 29^{\rm m} 11^{\rm s}$ &$+54^\circ 25'$ &
           0.715 & 26.7  &  0.4\rlap{$^\circ$} &32.1 &35.5 &1500 &920 \\
      B0450+55 &04 50 00  &+55 39  &0.341 & 14.6  &
      0.1 &\llap{$<$}8.5  & 8.7 &  59 & 110 \\
      B0655+64 &06 55 49  &+64 22  &0.196 &  8.8  &
      1.1 & 9.5 &14.0 &   5 &140 \\
      B0809+74 &08 09 03  &+74 38  &1.192 & 5.7  &
      0.3 &41.5 &33.9 &  79 &300 \\
      \smallskip
      B0823+26 &08 23 51  &+26 47  &0.531 & 19.4  &
      0.1 &39.9 &24.6 &  73 &410 \\
      B0834+06 &08 34 26  &+06 21  &1.274 & 12.8  &
      0.2 &44.3 &38.4 &  89 &1050 \\
      B0919+06 &09 19 35  &+06 51  &0.431 & 27.3  &
      0.7 &18.0 &15.1 &  52 &470 \\
      B0943+10 &09 43 27  &+10 06  &1.098 & 15.3  &
      0.0 &20.8 &17.0 &   4 &390 \\
      B0950+08 &09 50 31  &+08 10  &0.253 &  3.0  &
      1.9 &\llap{$<$}8.5  &13.8 & 400 &1500 \\
      \smallskip
      B1133+16 &11 33 27  &+16 08  &1.188 & 4.8 &
      1.0 &21.5 &22.1 & 257 &670 \\
      B1237+25 &12 37 12  &+25 10  &1.382 &  9.3  &
      0.2 &\llap{$<$}8.5   & 8.9 & 110 & 90 \\
      B1508+55 &15 08 03  &+55 43  &0.740 & 19.5 &
      0.6 &43.9 &49.4 & 114 &600 \\
      B1541+09 &15 41 14  &+09 39  &0.748 & 34.9  &
      0.1 &\llap{$<$}8.5   & 9.1 &  78 &290 \\
      B1642$-$03 &16 42 25  &$-$03 13  &0.388 & 35.7  &
      0.2 &12.5 &12.8 & 393 &510 \\
      \smallskip
      B1839+56 &18 39 51  &+56 38  &1.653 & 26.5  &
      1.0 & 9.7 &\llap{$<$}8.5   &  21 &110 \\
      B1919+21 &19 19 36  &+21 47  &1.337 & 12.4  &
      1.6 &22.3 &24.6 &  57 &1750 \\
      B2016+28 &20 16 00  &+28 31  &0.558 & 14.1  &
      0.9 & 9.7 &\llap{$<$}8.5   & 314 &400 \\
      B2110+27 &21 10 54  &+27 42  &1.203 & 24.7  &
      0.2 &\llap{$<$}8.5   & 8.5 &  18 & 90 \\
      B2217+47 &22 17 46  &+47 40  &0.539 & 43.5  &
      1.2 &37.4 &35.1 & 111 &1760 \\
      B2224+65 &22 24 17  &+65 20  &0.683 & 36.2  &
      0.5 &10.5 &13.2 &  22 &160 \\
      \hline
    \end{tabular}
  \end{center}
  \tablenotetext{\star}{Parameters in the first five columns and
  $S_{400}$ are from the catalog of Taylor {\em et al.} (1995; see
  also Taylor {\em et al.} 1993).  \nocite{tml93,tmlc95}}
\end{table}

\end{document}